\documentclass[11pt, titlepage]{article}
\usepackage{amsfonts}
\usepackage{amsmath}
\usepackage{amsthm}
\usepackage[font={small,it,stretch=1.2}]{caption}
\usepackage{changepage}
\usepackage{commath}
\usepackage{enumitem}
\usepackage[margin=2.5cm]{geometry}
\usepackage{graphicx}
\usepackage{hyperref}
\usepackage[utf8]{inputenc}
\usepackage{mathtools}
\usepackage[section]{placeins}                   
\usepackage{setspace}
\usepackage{subcaption}
\usepackage{titling}
\usepackage[dvipsnames]{xcolor}
\usepackage[british]{babel}
\usepackage{tabularx}
\usepackage{array}
\usepackage{longtable}

\usepackage[
  backend=biber,
  style=vancouver,
  citestyle=numeric-comp,
  sorting=none,           
  maxbibnames=3,
  minbibnames=3,
  natbib=true,            
  doi=true,
  isbn=false,
  url=false              
]{biblatex}
\let\cite\supercite
\let\citep\supercite
\let\citet\supercite
\let\citealp\supercite
\DeclareFieldFormat{journaltitle}{\textit{#1}}
\DeclareFieldFormat[book,mvbook,inbook]{title}{\textit{#1}}
\DeclareFieldFormat{booktitle}{\textit{#1}}
\DeclareFieldFormat{maintitle}{\textit{#1}}
\DeclareFieldFormat{pages}{\mkcomprange{#1}}
\DeclareFieldFormat{urldate}{\mkbibparens{#1, date last accessed}}

\DeclareFieldFormat[article,periodical]{volume}{\textbf{#1}}

\DefineBibliographyStrings{british}{%
  andothers = {\textit{et\addabbrvspace al\adddot}},
}

\hypersetup{
    colorlinks=TRUE,
    linktoc=all,
    linkcolor=violet,
    citecolor=teal
}
\newcommand{\midd}{\,|\,}

\newcommand{\indep}{\perp \!\!\! \perp}
\newcolumntype{Y}{>{\raggedright\arraybackslash}X}
\AtBeginDocument{
    \setlength\abovedisplayskip{-0.1cm}
    \setlength\abovedisplayshortskip{-0.1cm}
    \setlength\belowdisplayskip{0.4cm}
    \setlength\belowdisplayshortskip{0.4cm}
}

\allowdisplaybreaks

\begin{document}

\begin{titlepage}
    \centering
    \vspace*{2cm}
    
    {\LARGE\bfseries Repeated sampling of different individuals within the same clusters to improve precision of longitudinal estimators: the DISC design \par}
    \vspace{1.5cm}
    
    {\large Jordan Downey$^{1*}$, Avi Kenny$^{2,3}$\par}
    \vspace{1em}
    
    {\small
    $^1$Division of Biostatistics, University of California, Berkeley, CA, USA\\
    $^2$Department of Biostatistics and Bioinformatics, Duke University, Durham, NC, USA\\
    $^3$Global Health Institute, Duke University, Durham, NC, USA\par
    }

    \vspace{1em}
    
    {\small
    $^*$Corresponding author. Division of Biostatistics, University of California, Berkeley School of Public Health, 2121 Berkeley Way, Berkeley, CA 94704. Email: jordandowney@berkeley.edu
    }
    
    \vspace{2cm}
    {\large July 5, 2026\par}
    
    \vspace{1em}
    {\large \textbf{Word count:} 3246\par}
    
    \vfill
\end{titlepage}

\begin{center}
    \section*{Abstract}
\end{center}

\begin{adjustwidth}{0.5in}{0.5in}
    \textbf{Background:} Longitudinal studies often involve repeated cluster sampling of a population at multiple time points, such as in difference-in-differences (DID) studies. Although cohort designs typically lead to more efficient estimators relative to repeated cross-sectional (RCS) designs, they are often impractical. \\
    \textbf{Methods:} We describe the DISC (Different Individuals, Same Clusters) design, a sampling scheme that improves the precision of estimators in these settings. The DISC design represents a hybrid between a cohort and an RCS design, in which the researcher takes a single sample of clusters at baseline, but takes different cross-sectional samples of individuals within clusters at each time point. \\
    \textbf{Results:} We show analytically that the DISC design yields DID estimators with much higher precision relative to an RCS design, particularly if cluster effects are present. For example, for a design with two surveys, 40 clusters, and 25 individuals per cluster, the variance of a commonly-used DID treatment effect estimator is 2.3 times higher in the RCS design for an intraclass correlation coefficient (ICC) of 0.05 and 3.8 times higher for an ICC of 0.1. We also present results of a simulation study comparing the RCS and DISC designs, using both a simple DID estimator and a more complex doubly-robust DID (DRDID) estimator that leverages covariate information, and show gains in precision for both estimators when using the DISC design. Additionally, we illustrate DISC sampling using a household survey dataset from South Africa.\\
    \textbf{Conclusions:} Use of the DISC design can result in estimators that have substantially lower variance than the analogous estimator resulting from an RCS design. \\
    
    \noindent \textit{Keywords}: difference-in-differences, before and after survey, repeated cross-sectional sampling, cohort sampling \\

    \noindent{Key messages}: 
    \begin{itemize}
        \item The DISC (Different Individuals, Same Clusters) design is a practical hybrid between cohort and repeated cross-sectional (RCS) designs, in which the same clusters are sampled repeatedly, but within each cluster, different samples of individuals are collected at different time points.
        \item This design can be used in any longitudinal study that collects data through cluster sampling, including difference-in-differences designs, uncontrolled before and after studies, and interrupted time series designs, as well as in parallel cluster randomised trials and stepped wedge trials (where it is already commonly used).
        \item Estimators in a DISC design can have substantially lower variance relative to the corresponding estimators in a RCS design, especially in studies with small sample sizes and/or large cluster effects.
    \end{itemize}
    
\end{adjustwidth}

\section{Introduction}
Longitudinal studies involving repeated sampling of a population over time are ubiquitous in epidemiology, implementation science, global health, and economics \citep{diggle2002analysis}. A common quasi-experimental design used in this setting is difference-in-differences (DID) \citep{dimick2014methods}, which compares the change over time in a population receiving an intervention against the corresponding change in a control population. Under certain causal assumptions, DID produces a valid estimator of the intervention effect \citep{sant2020doubly}. 

Longitudinal designs, including DID, usually have either a cohort sampling structure (with the same individuals measured at multiple time points) or a repeated cross-sectional (RCS) structure (with independent samples drawn at different time points) \citep{feldman1994cohort,mann2003observational,hulley2013designing}. If an outcome of interest is positively correlated within individuals, a cohort design will typically lead to greater statistical efficiency relative to an RCS design \citep{diehr1995optimal,sant2020doubly}, due to the existence of cluster effects. However, cohort designs are often impractical. Reaching the same individuals over time can be logistically difficult, leading to high loss-to-follow-up \citep{diehr1995optimal} and potential selection bias if dropout is not at random, problems that are exacerbated in long-running studies \citep{feldman1994cohort}. Cohort designs are also not suitable in settings where individuals leave the risk set quickly, as in studies of neonatal or infant mortality.

A further complication, especially in studies of populations spread over large geographic areas, is that data are often collected through complex survey designs, such as multistage cluster sample household surveys \citep{lehtonen2004practical,chaudhuri2005survey}. These surveys may involve stratification, cluster sampling, unequal sampling weights, and other features which must be accounted for statistically during analysis \citep{rao1988resampling,lumley2004analysis,rabe2006multilevel}. The efficiency loss of an RCS-based DID estimator relative to a cohort-based estimator can be exacerbated under complex survey designs, particularly when cluster-level factors (measured or unmeasured) are strongly associated with the outcome. For example, a community's distance to the nearest health facility is associated with many health outcomes \citep{kenny2015remoteness}.

To address this, we propose a design in which the same clusters are sampled repeatedly, but different individuals are sampled within each cluster at different time points; we call this the \textit{DISC Design} (DISC = Different Individuals, Same Clusters). DISC can be used in any longitudinal study using cluster sampling, including DID designs, parallel cluster randomised trials (CRTs), stepped wedge trials \citep{hemming2015stepped}, uncontrolled before and after studies, and interrupted time series designs \citep{penfold2013use}. In this paper, we describe the DISC design and study its properties in the special case of DID studies, comparing DISC and RCS treatment effect estimators analytically, via statistical simulation, and in a case study using real-world data. This strategy is already commonly used in longitudinal CRTs, where it is (confusingly) called ``cross-sectional sampling''; for example, Lakshminarayan et al. repeatedly measured the same hospitals but different patients at different time points \cite{lakshminarayan2010cluster}. However, this strategy is far less common in quasi-experimental longitudinal designs, so it is useful to describe the design and define associated terminology to enable researchers to speak more precisely about potential design choices and trade-offs for a given study.

\section{Characterizing the DISC design}

\begin{figure}[ht!]
    \centering
    \includegraphics[width=0.9\linewidth]{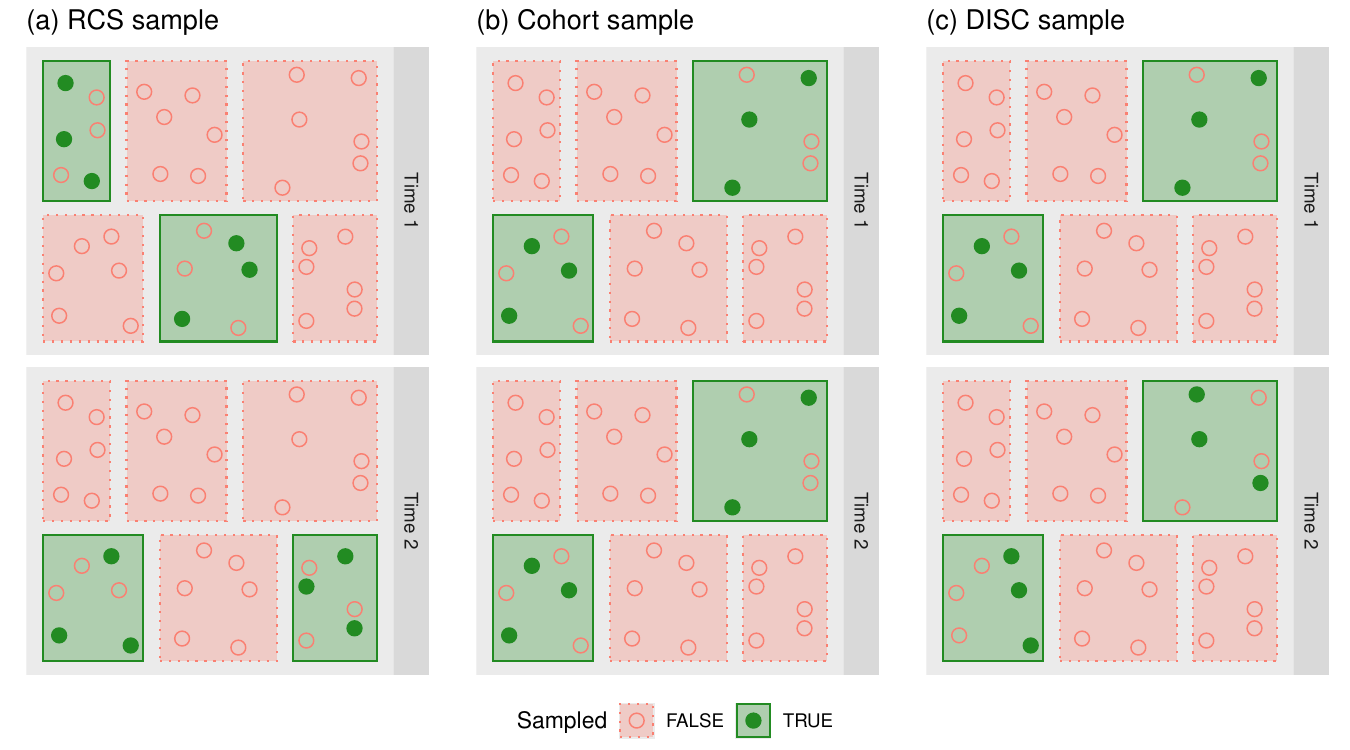}
    \captionsetup{width=0.9\linewidth}
    \caption{Visual representation of the DISC (Different Individuals, Same Clusters) design (panel (c)), compared to the repeated cross sectional (RCS; panel (a)) and cohort (panel (b)) sample designs. Rectangles represent clusters and circles represent individuals. Rectangles with solid borders and green shading represent those that are sampled and rectangles with dotted borders and pink shading represent those that are not sampled. Circles with a solid green fill represent those that are sampled and circles with a pink outline and no fill represent those that are not sampled. For each panel, the upper set of rectangles represents selections at the first time point and the lower set represents selections at the second time point.}
    \label{design_plot}
\end{figure}

Figure \ref{design_plot} illustrates differences between cohort, RCS, and DISC sampling. This figure visually describes a simple situation of two time points and one level of clustering, but the same principle generalises to higher dimensions. More than two levels of clustering is common in practice; for example, many household surveys involve first sampling a district or similar as the primary sampling unit, then sampling communities within districts, and finally sampling households within communities. In this scenario, a DISC design might involve selecting either the same communities or different communities at each time point. In a three-level cluster sampling structure, we propose a convenient notation to represent different designs: ``S-S-D'' (``S'' for ``same'', ``D'' for ``different'') or ``S-D-D'' to distinguish between DISC designs in which the samples at the second level of clustering are the same or different between surveys. The leftmost letter represents the primary (largest) sampling unit and the rightmost letter represents the smallest. For example, an ``S-S-D'' design would involve sampling the same district and the same community but different households. With this notation, ``S-S-S'' would represent a cohort study and ``D-D-D'' would represent a repeated cross-sectional study, and thus, this notation unifies all three sampling schemes. It can also be easily extended to designs with four or more sampling stages if needed. Similarly, the DISC design can be used when there are more than two time points, a situation sometimes referred to as a \textit{generalized difference-in-differences design} \citep{richardson2023generalized}.

\section{Analytic variance comparison between RCS and DISC designs}\label{sec_analytic}

Suppose that we are interested in a population containing $C$ clusters, where cluster $i$ contains $N_i$ individuals. At each time point $j \in \{1,2\}$, a sample of clusters is taken, and subsequently a sample of individuals is taken within each selected cluster. We assume that clusters are selected with probability-proportional-to-size (PPS) and that within each cluster, a sample of individuals of fixed size $m$ is taken \citep{rosen1997sampling}. For simplicity, assume the same number of clusters is sampled at both time points and that there is no missingness or non-response. Let $I_1 \subset \{1,2,...,C\}$ and $I_2 \subset \{1,2,...,C\}$ denote indexing sets corresponding to the clusters sampled at time points 1 and 2, respectively. For the DISC design $I_1=I_2$, and for the RCS design $I_1 \ne I_2$. For each sampled cluster $i$ and each time point $j$, let $K_{ij} \subset \{1,2,...,N_i\}$ be an indexing set (of size $m$) for the sample of individuals. Also, let $n$ be the total number of individuals sampled across all clusters at each time point.

Let $Y_{ijk}$ denote the outcome of interest corresponding to individual $k \in K_{ij}$ within cluster $i \in I_j$ at time point $j$, and let $X_i$ equal one if cluster $i$ is exposed to some ``treatment'' condition of interest at time 2 and zero otherwise (note that in a traditional DID design, no one is exposed to the treatment at time 1). For simplicity, we do not consider other covariates and we assume that an equal number of clusters are sampled from the ``treatment group'' (those with $X_i=1$) and the ``control group'' (those with $X_i=0$). In this simple setting, the participant-level DID estimand is given by

\begin{equation}\label{did_estimand}
    \delta \equiv
    \left\{ E(Y_{i2k} \midd X_i=1) - E(Y_{i1k} \midd X_i=1) \right\} -
    \left\{ E(Y_{i2k} \midd X_i=0) - E(Y_{i1k} \midd X_i=0) \right\} \,.
\end{equation}

\noindent That is, $\delta$ represents the difference between the change over time in the treatment group minus the change over time in the control group. If certain causal conditions are met \citep{sant2020doubly}, including the unconditional parallel trends assumption, the parameter $\delta$ can be identified with the average treatment effect on the treated (ATT). If weaker causal conditions (such as the \textit{conditional} parallel trends assumption) are met, the true ATT can be identified by a summary of the observed data distribution similar to \eqref{did_estimand}, but with each simple regression replaced by a G-computation parameter \citep{callaway2021difference} involving covariate information. In this section, we focus on the expression given in \eqref{did_estimand} for simplicity, but we consider the more complicated setting involving covariates in section \ref{sec_sims}.

The expression given in \eqref{did_estimand} suggests the estimator

\begin{align}\label{did_estimator}
\begin{split}
    \hat{\delta} &\equiv \left\{
        \frac{2}{n} \sum_{i\in I_2} \sum_{k\in K_{i2}} X_i Y_{i2k} -
        \frac{2}{n} \sum_{i\in I_1} \sum_{k\in K_{i1}} X_i Y_{i1k}
    \right\} \\ &\qquad - \left\{
        \frac{2}{n} \sum_{i\in I_2} \sum_{k\in K_{i2}} (1-X_i) Y_{i2k} -
        \frac{2}{n} \sum_{i\in I_1} \sum_{k\in K_{i1}} (1-X_i) Y_{i1k}
    \right\} \,,
\end{split}
\end{align}

\noindent which is equivalent to what one would obtain if a linear model including indicator variables for time 2, treatment group, and the interaction of treatment and time 2 (a common parameterisation of a simple DID linear model) was fit using generalised estimating equations with an independence working correlation matrix. Note that this corresponds to a participant-average treatment effect, as opposed to a group-average treatment effect \citep{imai2009essential,kahan2023estimands}, as all individuals are implicitly weighted equally (a consequence of PPS sampling; see, for example, \citealp{makela2018bayesian}).

We are interested in comparing $\text{Var}(\hat{\delta})$ between the two designs. To do so, we assume data at the population level were generated according to the following simple model:

\begin{equation}\label{eq_simple}
    Y_{ijk} = \mu_j + \delta X_{ij} + \alpha_i + \epsilon_{ijk} \,,
\end{equation}

\noindent where $\mu_j$ is the mean outcome at time $j$ (in the absence of treatment), $\delta$ is the treatment effect, $X_{ij} \equiv X_i I(j=2)$, $\epsilon_{ijk} \overset{iid}{\sim} N(0,\sigma^2)$, $\alpha_i \overset{iid}{\sim} N(0,\tau^2)$, and $\epsilon_{ijk} \indep \alpha_i$. We also denote $\hat{\delta}_\text{RCS}$ to equal $\hat{\delta}$ under the RCS design and $\hat{\delta}_\text{DISC}$ to equal $\hat{\delta}$ under the DISC design.

Under this model, it can be shown that under the RCS design, assuming the population of clusters is sufficiently large, the variance of $\hat{\delta}_\text{RCS}$ is given by

\begin{equation*}
    \text{Var}\left(\hat{\delta}_\text{RCS}\right) = 
    \frac{8 \left(m \tau^2 + \sigma^2 \right) }{n} \,.
\end{equation*}

\noindent Under the DISC design, the variance of $\hat{\delta}_\text{DISC}$ is given by

\begin{equation*}
    \text{Var}\left(\hat{\delta}_\text{DISC}\right) = 
    \frac{8\sigma^2}{n} \,.
\end{equation*}

\noindent See Appendix \ref{appx_var_calc} for a derivation of these results.

\begin{figure}[ht!]
    \centering
    \includegraphics[width=0.9\linewidth]{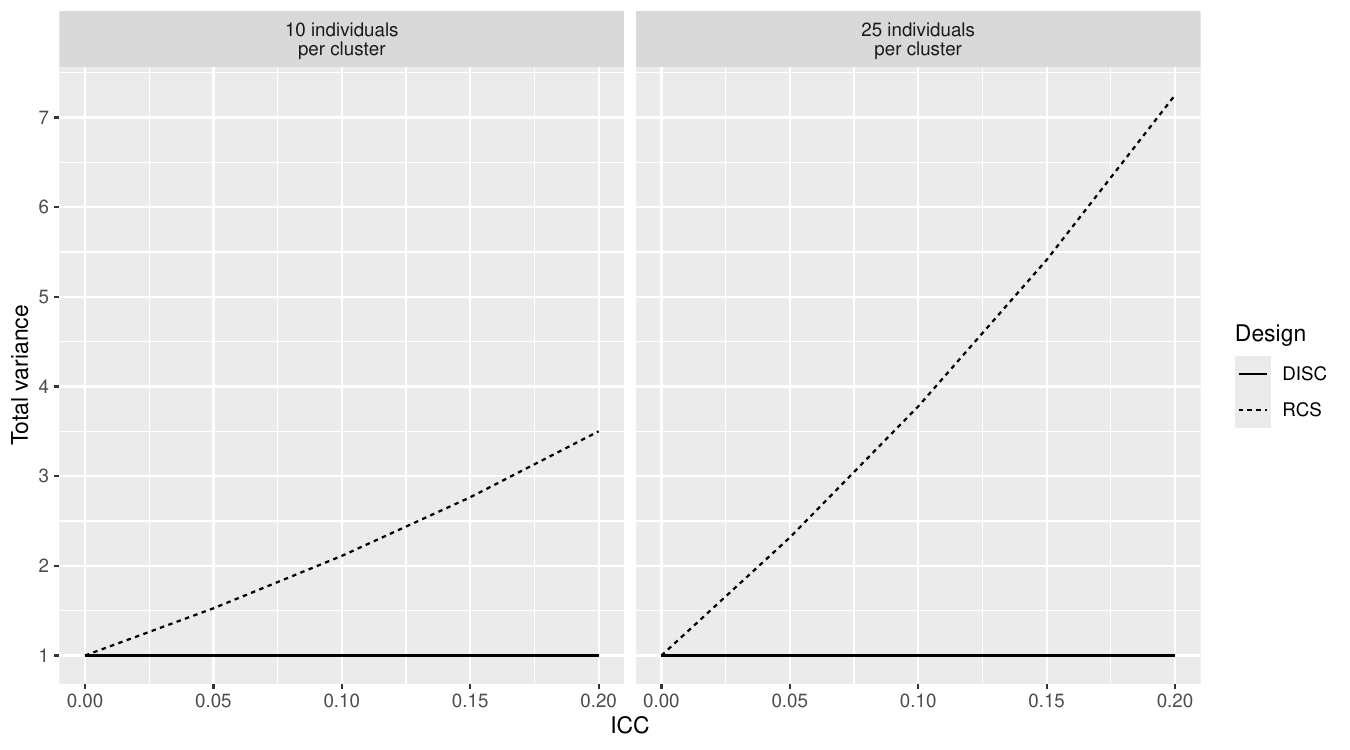}
    \captionsetup{width=0.9\linewidth}
    \caption{Total analytical variance comparing DISC (Different Individuals, Same Clusters) and RCS (repeated cross-sectional) designs as a function of the intraclass correlation coefficient (ICC), for a design involving a total of n=1,000 individuals. The panel on the left shows a design in which there are 100 clusters and 10 individuals sampled per cluster and the panel on the right shows a design in which there are 40 clusters and 25 individuals sampled per cluster.}
    \label{analytical_variance_figure}
\end{figure}

Considering the ratio of variances between the two designs can be informative. For a fixed sample size $n$, the variance of the treatment effect estimator in the RCS design relative to the DISC design is given by $1 + m\tau^2/\sigma^2$. It can also be helpful to write this ratio as a function of the intraclass correlation coefficient (ICC), defined as $\rho \equiv \tau^2 / (\tau^2+\sigma^2)$, a measure commonly used to quantify the proportion of the total variance of an outcome variable of interest due to cluster effects. The ICC is a common input of power formulas for CRTs \citep{hemming2020tutorial} as well as observational designs; \cite{korevaar2021intra} show that ICC values in the range of 0.02 to 0.1 are commonly encountered in CRTs, although this value is highly context-dependent. As a function of $\rho$, the ratio equals $1+m\rho/(1-\rho)$. In Figure \ref{analytical_variance_figure} we plot $\text{Var}\left(\hat{\delta}_\text{RCS}\right)$ and $\text{Var}\left(\hat{\delta}_\text{DISC}\right)$ as a function of $\rho$ for a fixed value of $n$ and several values of $m$; the residual standard deviation $\sigma^2$ is scaled such that $\text{Var}\left(\hat{\delta}_\text{DISC}\right)=1$, which implies that $\text{Var}\left(\hat{\delta}_\text{RCS}\right)$ is equivalent to the ratio of variances between the two designs.

In Figure \ref{analytical_variance_figure}, we see that, as expected, the variance in the DISC design does not depend on the ICC. Conversely, the variance in the RCS design can become extremely high relative to the variance in the DISC design for higher ICC values. For example, for a design in which 40 clusters and 25 individuals per cluster are sampled (for a total sample size of n=1,000), the variance of the treatment effect estimator is 2.3 times higher in the RCS design for an ICC of 0.05, 3.8 times higher for an ICC of 0.1, and 7.3 times higher for an ICC of 0.2.

\section{Simulation study}\label{sec_sims}

We conducted a simulation study to (1) confirm the analytic calculations presented in section \ref{sec_analytic}, and (2) to assess whether the gains in precision from using a DISC design hold when a more complicated estimator leveraging covariate information is used instead of the simple estimator studied in section \ref{sec_analytic}.

Data were generated to mimic a two-stage sampling design, with clusters sampled first and individuals second. A population of 1000 clusters was generated, half intervention and half control. A sample of 100 clusters was taken from this population using either the RCS design or the DISC design. In the RCS design, two different cluster samples (balanced between the intervention and control arms) were taken to represent the survey at time 1 and time 2, and in the DISC design, the same cluster sample was taken at both time points. Next, within each sampled cluster, a sample of 25 individuals was taken, with outcomes generated according to the equation

\begin{equation}\label{eq_covariates}
    Y_{ijk} = \mu_j + \delta X_{ij} + \beta^\top Z_{ijk} + \alpha_i + \epsilon_{ijk} \,,
\end{equation}

\noindent where $\alpha_i \overset{iid}{\sim}$ $N(0,\tau^2)$, $\epsilon_{ijk} \overset{iid}{\sim}$ $N(0,1)$, $(\mu_1,\mu_2,\delta)=(0,1,1)$. As in \eqref{eq_simple}, $\mu_1$ and $\mu_2$ represent the mean outcome values (in the absence of treatment) at times 1 and 2, respectively, and $\delta$ represents the treatment effect. Several values of $\tau$ were chosen to yield ICCs of 0.01, 0.05, 0.1, and 0.2, representing a reasonable range of ICC values encountered in practice based on the review of \cite{korevaar2021intra}. We ran simulations for cluster sizes of 10, 25, 50, and 100, also representing a reasonable range of cluster sizes based on \cite{korevaar2021intra}. $Z_{ijk}$ is a two-dimensional covariate vector with the first component sampled from a $\text{Bernoulli}(0.5)$ distribution and the second sampled from a $\text{Uniform}(0,1)$ distribution. In the first simulation, we set $\beta=(0,0)$ such that model \eqref{eq_covariates} is identical to model \eqref{eq_simple}, facilitating a comparison of the analytic variance formula to a simulation-based variance calculation. In the second simulation, we set $\beta=(0.5,10)$, such that the outcome is weakly associated with one covariate and strongly associated with a second covariate, to assess the performance of an estimator that leverages covariate information. Estimators considered included the linear estimator given by \eqref{did_estimator} and the doubly-robust difference-in-differences (DRDID) estimator studied in \cite{sant2020doubly}, which makes use of covariate information to gain precision. The DRDID estimator is a form of doubly-robust estimator, meaning that it is consistent for the true treatment effect if either the propensity score model or the outcome regression model (but not necessarily both) are correctly specified. Its specification is given in equation (3.1) of \cite{sant2020doubly}; we use the implementation in the \texttt{DRDID} package version 1.2.2, available on CRAN. All simulations were conducted in R version 4.3.2 and structured using the \texttt{SimEngine} package \citep{kenny2025simengine}. Results are based on 1,000 simulation replicates per scenario. Code to reproduce all simulation results is available at \url{https://github.com/jmdowney/disc}.

\begin{figure}[ht!]
    \centering
    \includegraphics[width=0.9\linewidth]{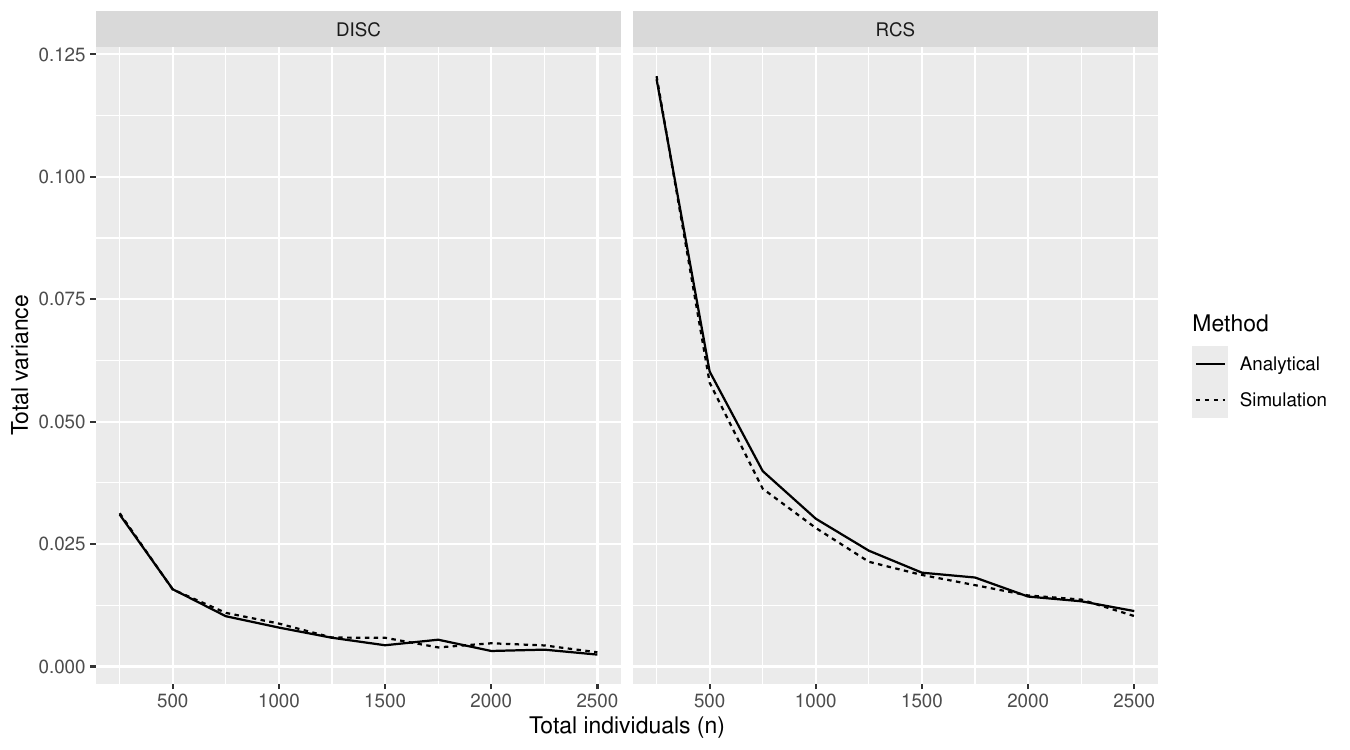}
    \captionsetup{width=0.9\linewidth}
    \caption{Theoretical variance and empirical (simulation-based) variance of the linear treatment effect estimator under DISC (Different Individuals, Same Clusters) and RCS (repeated cross-sectional) designs as a function of sample size. Data were generated without covariate effects and assuming 100 clusters sampled per time point and an ICC (intraclass correlation coefficient) of 0.1. Lines have been jittered slightly so that both lines can be seen.}
    \label{analytical_vs_simulation_figure_linear}
\end{figure}

Results from the first simulation are shown in Figure \ref{analytical_vs_simulation_figure_linear}. The simulation-based variance estimates (i.e., the variance of the treatment effect estimates across simulation replicates) match the theoretical variance for all sample sizes considered in both the RCS and the DISC design. Results for other choices of ICC and other cluster sizes similarly show alignment between theoretical and empirical variance and are given in Appendix \ref{appx_additional_sims}.

\begin{figure}[ht!]
    \centering
    \includegraphics[width=0.9\linewidth]{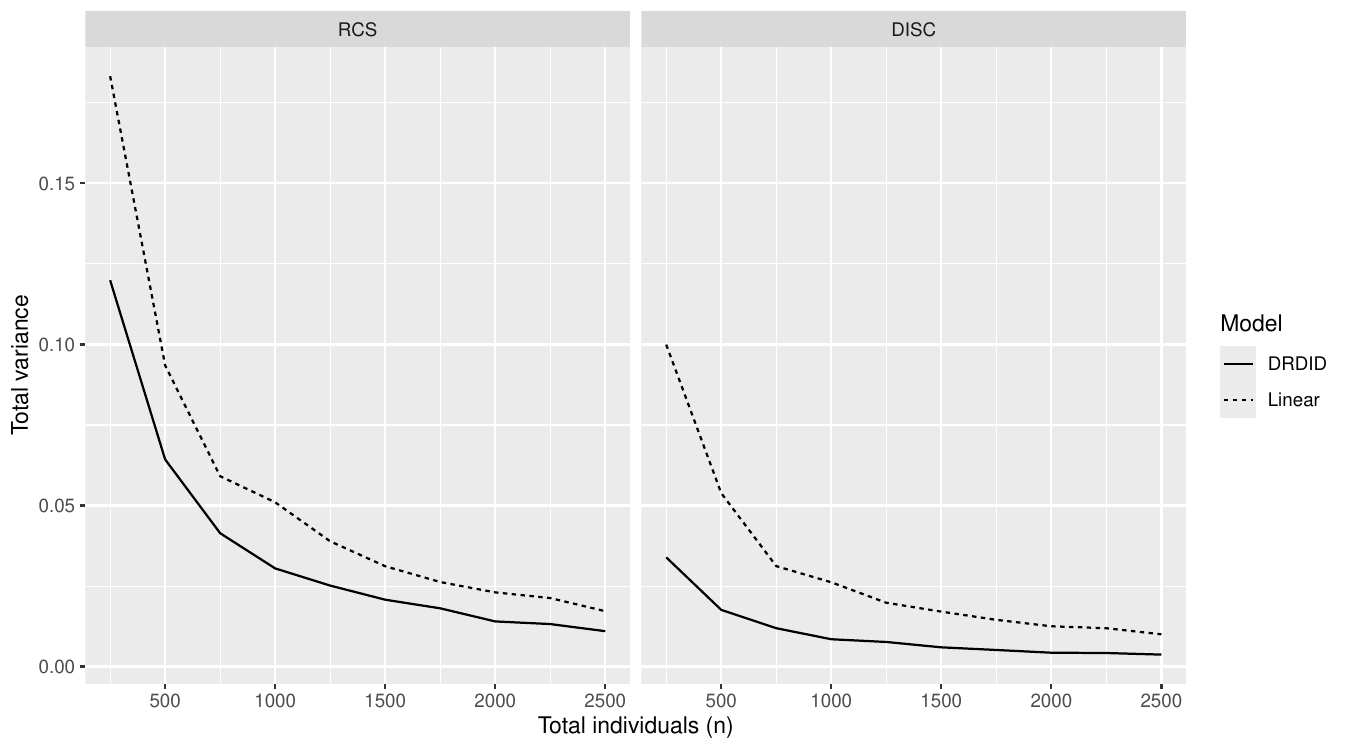}
    \captionsetup{width=0.9\linewidth}
    \caption{Estimated total variance under DISC (Different Individuals, Same Clusters) and RCS (repeated cross-sectional) designs, comparing a simple unadjusted linear model with a more complex adjusted DRDID (doubly robust difference-in-differences) model. The design assumes a uniform covariate, 100 clusters sampled, and an ICC (intraclass correlation coefficient) of 0.1.}
    \label{drdid_vs_linear_figure_2level}
\end{figure}

Results from the second simulation are shown in Figure \ref{drdid_vs_linear_figure_2level}. We observe an enormous gain in precision when using the DRDID estimator in the DISC design relative to the RCS design, confirming that the gain in precision is not restricted to the use of simple linear estimators but extends to settings in which covariate adjustment occurs and advanced doubly-robust methods are used. In both designs, the variance of the DRDID treatment effect estimator is lower than the variance of the linear estimator, which aligns with our expectations since both covariates adjusted for are associated with the outcome.

\section{Analysis of HDSS data}
To test the DISC design in a real-world scenario, we compared DISC and RCS sampling using household survey data from the Africa Health Research Institute (AHRI) Health and Demographic Surveillance System (HDSS) in KwaZulu-Natal, South Africa \citep{herbst2025ahri}, which aims to repeatedly measure an entire geographically bounded population. For exposition, we considered the set of 15,316 households appearing in both the 2013 and 2017 surveys to be the total population. We examined changes over time in several variables expected to have high ICCs and display results for the binary ``cattle ownership'' variable.

From this population, we mimicked 500 DISC samples and 500 RCS samples using the 50 ``local areas'' as clusters (each sample involved 10 clusters and 100 households per cluster). Within each sample, we calculated the change over time in the mean proportion of households that own cattle, with results shown in Figure 5.

\begin{figure}[ht!]
    \centering
    \includegraphics[width=0.9\linewidth]{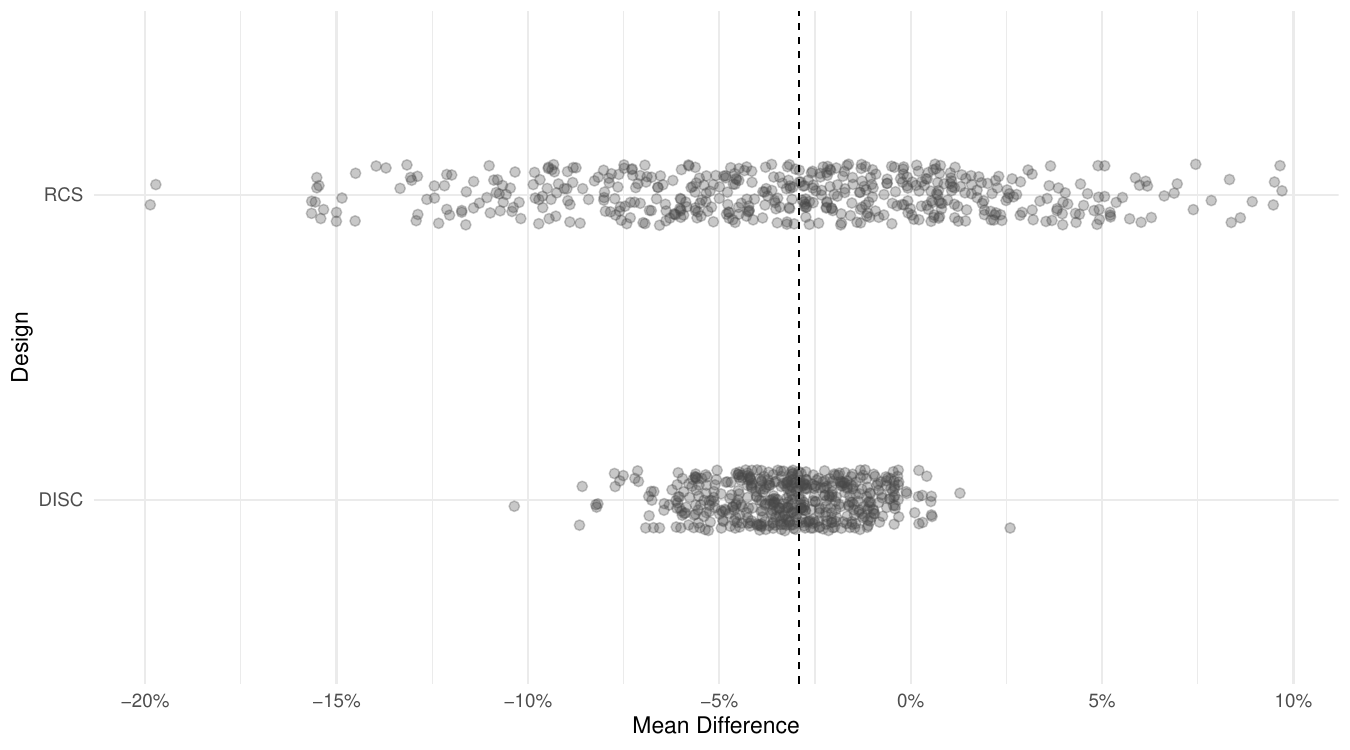}
    \captionsetup{width=0.9\linewidth}
    \caption{Change between 2013 and 2017 in the mean proportion of households who own cattle, from the Africa Health Research Institute Health and Demographic Surveillance System, under DISC (Different Individuals, Same Clusters) and RCS (repeated cross-sectional) sampling designs (with clusters sampled via probability proportional to size). Each data point represents an estimate from a single mimicked sample of 10 local areas (clusters) and 100 households. The vertical dotted line represents the population-level change in mean cattle ownership between 2013 and 2017. Points have been jittered vertically for visibility.}
    \label{2025-12-03_disc_rcs_ahri.pdf}
\end{figure}

While both sampling designs lead to unbiased estimates, the DISC design leads to far greater precision; the standard error of the RCS estimator is nearly 3 times higher than that of the DISC estimator.

\section{Discussion}

In this paper, we describe the DISC design (DISC = ``Different Individuals, Same Clusters''), a hybrid between a repeated cross-sectional (RCS) sampling design and a cohort sampling design for settings in which data are collected through cluster sampling. This sampling strategy can be used in any longitudinal study that utilises cluster sampling, including both experimental designs (CRTs, stepped wedge trials, etc.) and quasi-experimental designs (difference-in-differences, interrupted time series, uncontrolled before-and-after, etc.). Within the context of CRTs, the two-stage ``S-D'' design is already commonly used, where it is (confusingly) referred to as ``cross-sectional sampling''. However, we are unaware of any instances in which a multistage DISC design (such as the three-stage ``S-S-D'' design) has been used in a CRT. As shown in the simulations, meaningful improvements in variance can be made in a three-stage ``S-S-D'' design relative to a ``S-D-D'' design, and so this represents an opportunity to improve efficiency in CRTs. Focusing on the case of difference-in-differences (DID) treatment effect estimators, we show that use of the DISC design can result in estimators that have substantially lower variance than the analogous estimator resulting from an RCS design. The variance ratio increases as a function of the ICC, and for a fixed total sample size (in terms of number of individuals), it increases as the number of clusters decreases. For example, for a design with two surveys, 40 clusters, and 25 individuals sampled per cluster, the variance of a commonly-used DID treatment effect estimator is 2.3 times higher in the RCS design for an intraclass correlation coefficient (ICC) of 0.05 and 3.8 times higher for an ICC of 0.1. Since variance is inversely proportional to required sample size, this implies that a study powered at 90\% for detecting a treatment effect could be conducted with roughly half the sample size (for ICC=0.05) or one-quarter the sample size (for ICC=0.1) if a DISC design is used instead.

Although the analytical and simulation results presented in this paper focused on settings involving difference-in-differences estimators, we expect to see similar gains in precision for other types of studies. For example, a simple modification to the estimators and arguments given in section \ref{sec_analytic} can be done to show that the variance ratio is identical in the setting of uncontrolled before-and-after comparisons. Also, as mentioned in the Introduction, this sampling design is already implicitly used in the context of CRTs and stepped wedge trials.

In general, the choice of whether to use an RCS, DISC, or cohort design depends on a number of factors. To help practitioners decide which design is most appropriate given study objectives, logistical constraints, and population characteristics, a comparison of designs and tradeoffs is available in Table \ref{table_design_tradeoffs}. In practice, this decision is highly context-dependent, but in general, a DISC design is most appropriate when interest lies in estimating a treatment effect or change over time in a representative set of communities in a setting for which it is infeasible to track individuals over time (as in a cohort design) without high attrition. As shown in our simulations, the DISC design is particularly advantageous when large cluster-level effects are present. If instead researchers are interested in obtaining point estimates of quantities that are representative of the entire population at each time point, an RCS design will be more appropriate than a DISC design, especially in scenarios where cluster size or composition changes meaningfully over time, such as when there is significant migration. In these scenarios, a DISC or cohort sample at the second time point may no longer be representative of the population. However, if a change in an outcome over time is observed in a setting where there are ongoing shifts in the underlying population, this change can be partitioned into the portion resulting from changes in the outcome among the people who were in the population for the duration of the study versus the portion resulting from people entering or leaving the study population; if interest lies in changes within the subpopulation who remained in the study, the RCS design will be inferior.

Real-world challenges associated with implementing a DISC design include attrition of clusters between baseline and endline, changing cluster composition over time, and time-varying cluster-level effects. While cluster attrition will only occur in specific situations, if it is expected, practitioners can consider building this into power calculations and accounting for cluster dropout in the analysis plans using missing data methods. If the composition of clusters is expected to change over time in a setting where a treatment effect is being estimated, measuring and controlling for relevant individual-level confounders can help guard against bias. Similarly, if time-varying cluster-level effects are present and measurable, they can be controlled for in the analysis.

\begin{table}[ht!]
\footnotesize
\setlength{\tabcolsep}{4pt}       
\renewcommand{\arraystretch}{1.1}
\begin{tabularx}{\textwidth}{|Y|Y|Y|Y|}
\cline{1-4}
 & \textbf{Cohort Design} & \textbf{DISC (Different Individuals, Same Clusters) Design} & \textbf{RCS (repeated cross-sectional) Design} \\
\cline{1-4}
\textbf{Study objective} & Estimate changes over time or treatment effects in a fixed, representative set of individuals & Estimate changes over time or treatment effects in a fixed, representative set of communities & Estimate changes over time in a population, including changes due to demographic shifts and migration, with estimates at each time point representative of the overall population \\
\cline{1-4}
\textbf{Statistical power \& efficiency} & Highest efficiency; estimators remove both cluster-level and individual-level variance components by comparing individuals to themselves over time & Medium efficiency; estimators remove cluster-level variance components by comparing communities to themselves over time & Lowest efficiency; estimators include cluster-level and individual-level variance contributions \\
\cline{1-4}
\textbf{Sample size} & Requires the smallest number of clusters and/or individuals per cluster & Required sample size is lower than the RCS design and higher than the cohort design & Requires the largest number of clusters and/or individuals per cluster \\
\cline{1-4}
\textbf{Feasibility} & Feasible only when high retention and robust individual-level tracking over time is possible & Much more feasible than a cohort design when it is difficult to locate the same individuals over time; individual dropout is not a concern & Feasibility is similar to the DISC design, but a larger sample size will be required \\
\cline{1-4}
\textbf{Representativeness} & Sample is representative of the population at baseline but may not be at later time points if there are demographic shifts within or between communities; attrition can leads to bias over time & Sample is representative of the population at baseline but may not be at subsequent time points if demographic shifts within sampled communities differ from those within non-sampled communities; no bias from attrition & Sample is representative of the population at baseline and at all subsequent time points; no bias from attrition \\
\cline{1-4}
\textbf{Intraclass correlation coefficient (ICC) considerations} & Largest efficiency gains occur with high ICC & Largest efficiency gains occur with high ICC & Poor efficiency with high ICC \\
\cline{1-4}
\end{tabularx}
\captionsetup{width=0.9\linewidth,skip=2em}
\caption{Considerations when choosing between cluster sampling study designs}
\label{table_design_tradeoffs}
\end{table}

One possible direction for future research would be to compare DISC against strategies that involve adjustment for cluster-level covariates in the analysis phase. Of course, any such strategy would only work if the cluster-level covariates are measurable. A second research direction would be to consider a DISC variant for open cohort longitudinal designs \citep{kasza2020sample}, in which individuals may enter or exit the cohort during the study period; depending on the mechanism through which individuals leave the cohort, it may be possible to create a design involving a hybrid of cohort sampling (for individuals who remain in the cohort) and DISC sampling (to replace individuals who leave the cohort with individuals who enter). A third direction would be to examine the impact of varying the cluster autocorrelation coefficient \citep{hooper2015cluster, teerenstra2012simple} on results. When cluster autocorrelation coefficient is high (i.e., when effects are at the level of the cluster-time interaction rather than the level of the cluster), it would be expected that precision gains from the DISC design would be attenuated, which is important to explore and quantify.

\appendix

\section*{Declarations}
\subsection*{Author contributions}
A.K. conceived the method, developed the theoretical framework, and supported analytic calculations. J.D. derived analytic results, programmed the simulations, and analyzed HDSS data. Both authors drafted and revised the manuscript and approve of the final version.
\subsection*{Supplementary data}
Supplementary data is available at \textit{IJE} online.
\subsection*{Conflict of interest}
None declared.
\subsection*{Funding}
None.
\subsection*{Data availability} 
All simulation code and data can be found at \url{https://github.com/jmdowney/disc}. HDSS data can be requested through the official AHRI Data Repository found at \url{https://data.ahri.org/index.php/home}.
\subsection*{Use of artificial intelligence (AI) tools}
AI tools were not used in the preparation of this manuscript.

\section*{References}
\printbibliography[heading=none]

\section{Variance calculations}\label{appx_var_calc}

The estimator given in \eqref{did_estimator} is defined as

\begin{align*}
\begin{split}
    \hat{\delta} &\equiv \left\{
        \frac{2}{n} \sum_{i\in I_2} \sum_{k\in K_{i2}} X_i Y_{i2k} -
        \frac{2}{n} \sum_{i\in I_1} \sum_{k\in K_{i1}} X_i Y_{i1k}
    \right\} \\ &\qquad - \left\{
        \frac{2}{n} \sum_{i\in I_2} \sum_{k\in K_{i2}} (1-X_i) Y_{i2k} -
        \frac{2}{n} \sum_{i\in I_1} \sum_{k\in K_{i1}} (1-X_i) Y_{i1k}
    \right\} \,,
\end{split}
\end{align*}

Assume that the population of clusters is sufficiently large such that the probability of selecting the same cluster at both time points is zero (a simplifying assumption to ease calculations). Then under the RCS design, labelling our estimator as $\hat{\delta}_\text{RCS}$, it holds that

\begin{align*}
    \text{Var} \left( \hat{\delta}_\text{RCS} \right)
    &= \text{Var} \Bigg[ \Bigg\{
        \frac{2}{n} \sum_{i\in I_2} \sum_{k\in K_{i2}} X_i Y_{i2k} -
        \frac{2}{n} \sum_{i\in I_1} \sum_{k\in K_{i1}} X_i Y_{i1k}
    \Bigg\} \\ &\qquad\qquad - \Bigg\{
        \frac{2}{n} \sum_{i\in I_2} \sum_{k\in K_{i2}} (1-X_i) Y_{i2k} -
        \frac{2}{n} \sum_{i\in I_1} \sum_{k\in K_{i1}} (1-X_i) Y_{i1k}
    \Bigg\} \Bigg] \\
    &= 2\, \text{Var} \left\{
        \frac{2}{n} \sum_{i\in I_2} \sum_{k\in K_{i2}} X_i Y_{i2k} -
        \frac{2}{n} \sum_{i\in I_2} \sum_{k\in K_{i2}} (1-X_i) Y_{i2k}
    \right\} \\
    &= 2\, \text{Var} \left\{
        \frac{2}{n} \sum_{i\in I_2} \sum_{k\in K_{i2}} (2X_i-1) Y_{i2k}
    \right\} \\
    &= \frac{8}{n^2} \left[
        \sum_{i\in I_2} \left\{ (2X_i-1)^2 \text{Var} \sum_{k\in K_{i2}} Y_{i2k} \right\}
    \right] \\
    &= \frac{8}{n^2} \left[
        \sum_{i\in I_2} \left\{
            \text{Var} \sum_{k\in K_{i2}} (\mu_2 + \delta X_{ij} + \alpha_i + \epsilon_{ijk})
        \right\}
    \right] \\
    &= \frac{8}{n^2} \left[
        \sum_{i\in I_2} \left\{
            \text{Var} (m \alpha_i) +
            \sum_{k\in K_{i2}} \text{Var} (\epsilon_{ijk})
        \right\}
    \right] \\
    &= \frac{8m}{n^2} \left[
        \sum_{i\in I_2} \left\{
            m \tau^2 +
            \sigma^2
        \right\}
    \right] \\
    &= \frac{8 \left(m \tau^2 + \sigma^2 \right) }{n} \,
\end{align*}

Under the DISC design, we can label this estimator as $\hat{\delta}_\text{DISC}$, and making a similar simplifying assumption that the probability of sampling the same individual at both time points is zero, it holds that

\begin{align*}
    \text{Var} \left( \hat{\delta}_\text{DISC} \right)
    &= \text{Var} \Bigg[ \Bigg\{
        \frac{2}{n} \sum_{i\in I_2} \sum_{k\in K_{i2}} X_i Y_{i2k} -
        \frac{2}{n} \sum_{i\in I_1} \sum_{k\in K_{i1}} X_i Y_{i1k}
    \Bigg\} \\ &\qquad\qquad - \Bigg\{
        \frac{2}{n} \sum_{i\in I_2} \sum_{k\in K_{i2}} (1-X_i) Y_{i2k} -
        \frac{2}{n} \sum_{i\in I_1} \sum_{k\in K_{i1}} (1-X_i) Y_{i1k}
    \Bigg\} \Bigg] \\
    &= \frac{4}{n^2} \text{Var} \Bigg[
        \sum_{i\in I_1} X_i \Bigg( \sum_{k\in K_{i2}} Y_{i2k}
        - \sum_{k\in K_{i1}} Y_{i1k} \Bigg)
    \\ &\qquad\qquad\qquad -
        \sum_{i\in I_1} (1-X_i) \Bigg( \sum_{k\in K_{i2}} Y_{i2k}
        - \sum_{k\in K_{i1}} Y_{i1k} \Bigg)
    \Bigg] \\
    &= \frac{4}{n^2} \text{Var} \Bigg[
        \sum_{i\in I_1} X_i \Bigg( \sum_{k\in K_{i2}} (\delta + \epsilon_{i2k})
        - \sum_{k\in K_{i1}} \epsilon_{i1k} \Bigg)
    \\ &\qquad\qquad\qquad -
        \sum_{i\in I_1} (1-X_i) \Bigg( \sum_{k\in K_{i2}} \epsilon_{i2k}
        - \sum_{k\in K_{i1}} \epsilon_{i1k} \Bigg)
    \Bigg] \\
    &= \frac{4}{n^2} \text{Var}
        \sum_{i\in I_1} (2X_i-1) \Bigg( \sum_{k\in K_{i2}} \epsilon_{i2k}
        - \sum_{k\in K_{i1}} \epsilon_{i1k} \Bigg) \\
    &= \frac{4}{n^2}
        \sum_{i\in I_1} \Bigg( \sum_{k\in K_{i2}} \sigma^2
        + \sum_{k\in K_{i1}} \sigma^2 \Bigg) \\
    &= \frac{8\sigma^2}{n} \,
\end{align*}

\section{Additional simulation results}\label{appx_additional_sims}

\begin{figure}[ht!]
    \centering
    \includegraphics[width=0.89\linewidth]{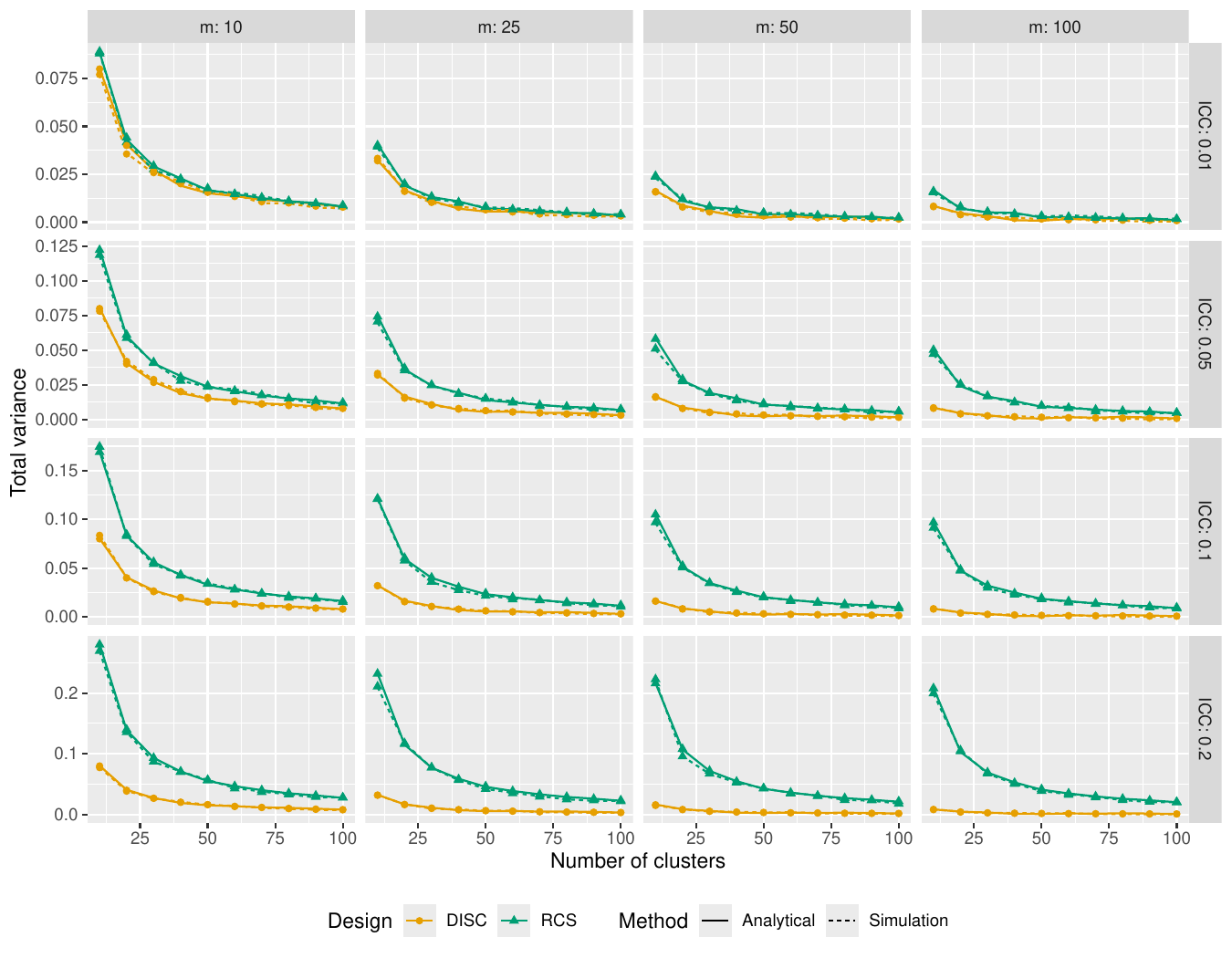}
    \captionsetup{width=0.89\linewidth}
    \caption{Estimated total variance of the linear treatment effect estimator as a function of the number of clusters under DISC (Different Individuals, Same Clusters) and RCS (repeated cross-sectional) designs. Dotted lines represent simulation-based variance and solid lines represent theoretical analytic variance. Green lines with triangle points represent RCS designs and orange lines with circular points represent DISC designs. The column facets correspond to four different cluster sizes (10, 25, 50, and 100) and the row facets correspond to four different ICC (intraclass correlation coefficient) values (0.01, 0.05, 0.1, 0.2).}
    \label{supp_fig_1}
\end{figure}

In this section, we display additional simulation results that examine sensitivity of results to changes in the data-generating mechanism. Figure \ref{supp_fig_1} shows complete simulation results comparing the empirical variance of the linear estimator to the theoretical variance under the DISC design versus the RCS design. Results are shown for four cluster sizes (10, 25, 50, and 100) and four ICC values (0.01, 0.05, 0.1, 0.2). These results confirm that the simulation-based empirical variance matches the analytical variance across a variety of data-generating mechanisms.

\begin{figure}[ht!]
    \centering
    \includegraphics[width=0.9\linewidth]{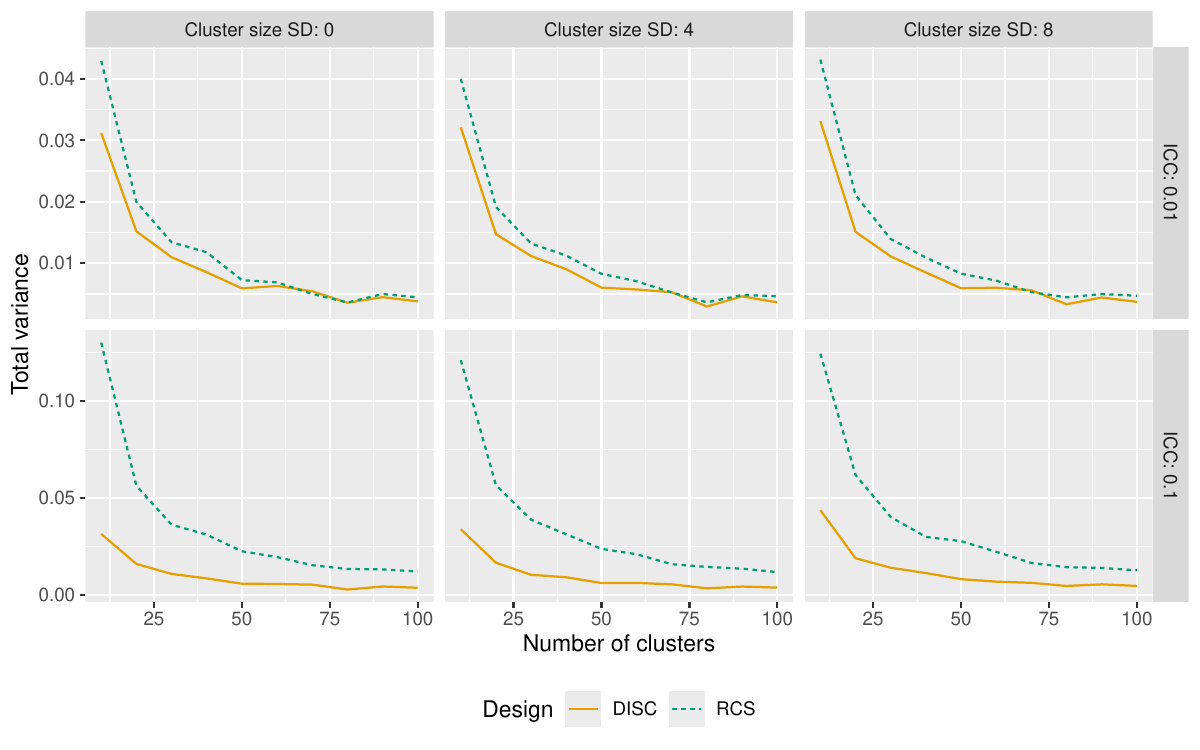}
    \captionsetup{width=0.9\linewidth}
    \caption{Total variance of the linear treatment effect estimator under DISC (Different Individuals, Same Clusters) and RCS (repeated cross-sectional) designs, estimated via simulation, as a function of the number of clusters. Dotted green lines represent RCS designs and solid orange lines represent DISC designs. The column facets correspond to three different cluster size standard deviations (0, 4, and 8) and the row facets correspond to two different ICC (intraclass correlation coefficient) values (0.01, and 0.1).}
    \label{supp_fig_2}
\end{figure}

Figure \ref{supp_fig_2} shows how the variance of the linear estimator changes as we increase variation in cluster sizes. The same simulation setup is used, except cluster sizes were generated by sampling from a Normal distribution with mean 25 and a standard deviation of either 0 (implying no variation in cluster sizes), 4, or 8. Results are shown for two ICC values (0.01 and 0.1). These results show that the benefits of the DISC design relative to the RCS design persist even when cluster sizes are highly variable.

\begin{figure}[ht!]
    \centering
    \includegraphics[width=0.9\linewidth]{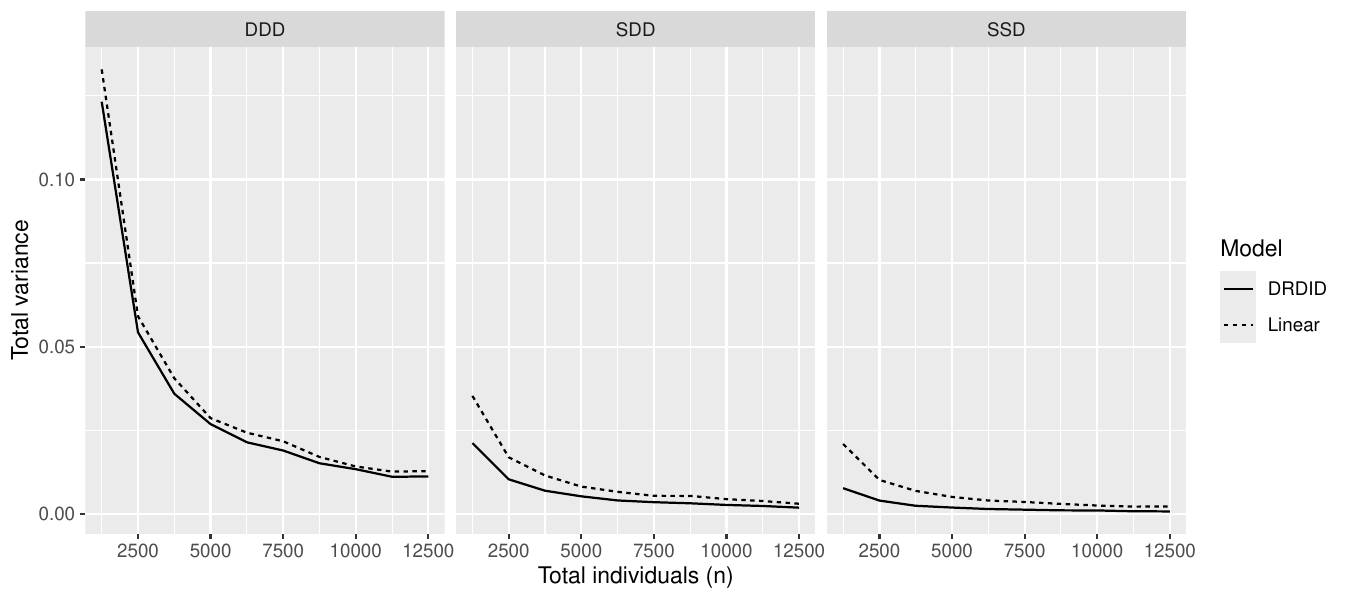}
    \captionsetup{width=0.9\linewidth}
    \caption{Estimated total variance under three different three-level sampling designs (Different-Different-Different, or D-D-D; Same-Different-Different, or S-D-D; and Same-Same-Different, or S-S-D), comparing a simple unadjusted linear model (dotted line) with a more complex adjusted doubly robust difference-in-differences model (solid line). Data are generated with a standard uniform covariate, 100 clusters, and an ICC (intraclass correlation coefficient) of 0.1.}
    \label{drdid_vs_linear_figure_3level}
\end{figure}

Figure \ref{drdid_vs_linear_figure_3level} shows the total variance of both the linear and DRDID estimators under a three-level sampling design. Data were generated according to a three-level hierarchical model, with random intercepts at the first and second levels. Other than the additional random intercept at the second level and the sampling mechanism, the data-generating mechanism is identical to the one used for Figure \ref{drdid_vs_linear_figure_2level}. As expected, this figure shows that both versions of the DISC design (S-D-D and S-S-D) are superior to the RCS design (D-D-D) in terms of estimated variance. Furthermore, and also as expected, we see that variance is lower for the S-S-D design relative to the S-D-D design.

Finally, we give additional simulation results in Table \ref{table_sim_results}, including estimated bias, variance, mean squared error, and 95\% confidence interval coverage. These correspond to the simulation results displayed in Figure \ref{analytical_vs_simulation_figure_linear}, with modifications in terms of the ICC, number of clusters, and the cluster sizes. The main purpose of displaying these results is to show that theoretical expectations match simulation results; we can see that the linear estimator is unbiased across all scenarios, that variance is roughly equal to MSE, and that nominal 95\% confidence interval coverage is maintained.

\begin{table}[ht!]
\begin{tabular}{llllllll}
\cline{1-8}
\textbf{Design} & \textbf{ICC} & \textbf{Number of clusters} & \textbf{Cluster size} & \textbf{Bias} & \textbf{Variance} & \textbf{MSE} & \textbf{Coverage} \\
\cline{1-8}
DISC & 0.01 & 20 & 25 & 0.007 & 0.016 & 0.016 & 94.3\% \\
DISC & 0.01 & 20 & 100 & 0.006 & 0.006 & 0.006 & 95.0\% \\
DISC & 0.01 & 50 & 25 & 0.001 & 0.004 & 0.004 & 96.1\% \\
DISC & 0.01 & 50 & 100 & 0.001 & 0.002 & 0.002 & 95.4\% \\

DISC & 0.1 & 20 & 25 & 0.001 & 0.015 & 0.015 & 95.6\% \\
DISC & 0.1 & 20 & 100 & -0.005 & 0.006 & 0.006 & 95.4\% \\
DISC & 0.1 & 50 & 25 & 0.000 & 0.004 & 0.004 & 95.0\% \\
DISC & 0.1 & 50 & 100 & -0.001 & 0.002 & 0.002 & 96.0\% \\

RCS & 0.01 & 20 & 25 & -0.002 & 0.020 & 0.020 & 95.4\% \\
RCS & 0.01 & 20 & 100 & -0.003 & 0.008 & 0.008 & 95.6\% \\
RCS & 0.01 & 50 & 25 & -0.004 & 0.008 & 0.008 & 95.1\% \\
RCS & 0.01 & 50 & 100 & -0.002 & 0.003 & 0.003 & 96.6\% \\

RCS & 0.1 & 20 & 25 & -0.006 & 0.058 & 0.058 & 95.1\% \\
RCS & 0.1 & 20 & 100 & 0.008 & 0.022 & 0.022 & 96.8\% \\
RCS & 0.1 & 50 & 25 & 0.000 & 0.048 & 0.047 & 95.0\% \\
RCS & 0.1 & 50 & 100 & -0.002 & 0.018 & 0.018 & 94.9\% \\

\cline{1-8}
\end{tabular}
\captionsetup{width=0.9\linewidth,skip=2em}
\caption{Simulation results for a subset of data-generating parameters across 1,000 simulation replicates, including bias, variance, mean squared error, and 95\% confidence interval coverage.}
\label{table_sim_results}
\end{table}


\end{document}